# The Venus Life Equation

**Hypothesis Paper, Submitted to Astrobiology**

**Noam R. Izenberg\*** *([noam.izenberg@jhuapl.edu](mailto:noam.izenberg@jhuapl.edu), 443-778-7918, Johns Hopkins University Applied Physics Laboratory (JHUAPL), Laurel, Maryland, USA*. ORCID: [0000-0003-1629-6478](https://orcid.org/0000-0003-1629-6478))

**Co-Authors**

**D. M. Gentry**, [diana.gentry@nasa.gov](mailto:diana.gentry@nasa.gov), NASA Ames Research Center, Moffett Field, CA, USA

**D J. Smith**, [david.j.smith-3@nasa.gov](mailto:david.j.smith-3@nasa.gov), NASA Ames Research Center, Moffett Field, CA, USA

**M. S. Gilmore**, [mgilmore@wesleyan.edu](mailto:mgilmore@wesleyan.edu), Wesleyan University, Middletown, CT, USA

**D. H. Grinspoon**, [grinspoon@psi.edu](mailto:grinspoon@psi.edu), Planetary Science Institute, Washington, DC, USA

**M. A. Bullock**, [bullock@stcnet.com](mailto:bullock@stcnet.com), Science and Technology Corp., Boulder, CO, USA

**P. J. Boston**, [penelope.j.boston@nasa.gov](mailto:penelope.j.boston@nasa.gov), NASA Ames Research Center, Moffett Field, CA, USA

**G. P. Słowik**, [grzegslowik@o2.pl](mailto:grzegslowik@o2.pl), Institute of Materials and Biomedical Engineering, Faculty of Mechanical Engineering, University of Zielona Góra, Poland

\*Corresponding author.

**Abstract**

Ancient Venus and Earth may have been similar in crucial ways for the development of life, such as liquid water oceans, land-ocean interfaces, favorable chemical ingredients and energy pathways. If life ever developed on, or was transported to, early Venus from elsewhere, it might have thrived, expanded and then survived the changes that have led to an inhospitable surface on Venus today. The Venus cloud layer may provide a refugium for extant life that persisted from an earlier more habitable surface environment. We introduce the Venus Life Equation - a theory and evidence-based approach to calculate the probability of extant life on Venus, $L$, using three primary factors of life: Origination, Robustness, and Continuity, or $L = O \cdot R \cdot C$. We evaluate each of these factors using our current understanding of Earth and Venus environmental conditions from the Archaean to the present. We find that the probability of origination of life on Venus would be similar to that of the Earth and argue that the other factors should be nonzero, comparable to other promising astrobiological targets in the solar system. The Venus Life Equation also identifies poorly understood aspects of Venus that can be addressed by direct observations with future exploration missions.

**Introduction:** One of the biggest motivators for exploring the solar system beyond Earth is to determine whether extant life currently exists, or now-extinct life once existed, on worlds beyond ours. Current knowledge about the past and present climate of Venus suggests it once had an extended period – perhaps 2 billion years – where a water ocean and land-ocean interfaces could have existed on the surface, under conditions resembling those of Archaean Earth (Way *et al.*, 2016; Way and Del Genio, 2020). Although today the Venus surface (450 °C, 92 bars) is not hospitable to life as we know it, there is a zone of the Venus middle atmosphere, at around 55 km altitude just above the sulfuric acid middle cloud layer, where some conditions are Earth-like (Figure 1) (Cavicchioli, 2002; Titov *et al.*, 2007). Further interest stems from energetic and chemical observations, not explained by current models, that have similarities to known biological phenomena: the presence of phosphine (Greaves *et al.*, 2020) and regions of strong UV absorption (Limaye *et al.*, 2018). The question of whether life could have – or could still – exist on the Earth's closest neighbor is more open today than it's ever been (Morowitz and Sagan, 1967; Limaye *et al.*, 2018; Seager *et al.*, 2020).

This paper approaches the question of extant life on Venus in a similar manner to the strategy of the Drake Equation established for estimating probabilities of extraterrestrial intelligent life (Drake, 1965; Burchel, 2006). We approach the question of whether life exists currently "on" Venus (we include the planet's atmosphere in this definition) as an exercise in informal probability – seeking the qualitative likelihood (i.e. *not* the statistical likelihood function) of the answer being nonzero. The fundamental goal of the Venus Life Equation (VLE) is to provide a scaffolding for estimating the chance of extant life based on factors that can be constrained or quantified through observation, experiment, and data-based modeling. Here we offer a framework in which these estimations can be made.

**The Venus Life Equation:** The Venus Life Equation (Figure 2) is expressed as:

$$L = O \bullet R \bullet C \tag{1}$$

where $L$ is the likelihood (zero to 1 representing "no chance", to "a certainty") of there being current Venus life, $O$ (origination) is the chance life ever began and became established on Venus, $R$ (robustness) is the potential size and diversity of the Venus biosphere past or present, and $C$ (continuity) is the chance that conditions amenable to life persisted spatially and temporally until now. Subfactors are similarly treated as likelihoods (zero to 1).

The VLE is intentionally agnostic about the scale and type of life that might be present on Venus. Simple or complex, microbial or multicellular; any or all (e.g., Ganti, 2003) would mean $L > 0$. Our focus in this paper is applying this systematic framework to evaluate the current dominant hypothesis for extant Venus life: a global, persistent airborne ecosystem consisting of organisms small enough to be suspended in or among the cloud aerosols (Figure 3). Therefore, we examine the constraints and unknowns relevant to the global history and current state of the Venus clouds. However, the same framework could be applied to develop a plausible range for the likelihood of past Venus life, and/or of life in a different niche. Other suggestions that have been put forward for Venus include surface life adapted to use supercritical carbon dioxide as a solvent (Budisa and Schulze-Makuch, 2014), and subsurface microbes in refugia of highly pressurized water (Schulze-Makuch *et al.*, 2005).

The following sections describe each variable of the VLE. The purposefully global scale of Eq. 1, and the major subfactors exemplified by Eqs. 2 and 3, are intended to present an argument for pursuing deeper investigations in those high-level areas (i.e., prioritizing future research efforts). Subvalues of O, R, and C could be broken out to far finer levels of detail than is possible to cover in a single paper - eventually to scales (Figure 4) appropriate to specific scientific investigations in the lab, analogue field sites on Earth, or in-situ measurements on Venus.

**Origination**: Life on a planet can start via independent abiogenesis, or importation from elsewhere (panspermia), where:

$$O = 1 - ((1 - O_A) \bullet (1 - O_P)) = O_A + O_P - (O_A \bullet O_P) \qquad (2)$$

where $O_A$ is the likelihood of origin by *abiogenesis* and $O_P$ is the likelihood of origin by *panspermia*. The final term in Eqn. 2 removes a potential double-counting if life has arisen by both abiogenesis and panspermia, i.e., the probability of two separate geneses. If $O_A$ and $O_P$ are both presumed to be small, this term is negligible, leaving O as a simple sum. There are other methods of estimating Origination (e.g., Damer and Deamer, 2020) that delve into greater detail or use more or different subfactors. For the VLE exercise, we approach the question with the most general possible framework.

$O_A$ depends on how likely it is for life to arise independently. In our own solar system, empirically, we assign $O_A \sim 1$ for Earth. For other bodies, considering the lack of other definitive evidence, we assume the baseline is effectively zero, with probabilities increasing if current or historical conditions become similar to those of early Earth. Though much remains unknown about Venus's past history, it is clear that its early state, modeled to have clement oceans for 2-3 Ga (Way *et al.*, 2016; Way and DelGenio, 2020), was very different from its current state, in which liquid water cannot exist on the surface. If we assume that early Venus followed an environmental trajectory similar to Earth's for ≥2 Ga years, the upper bound of $O_A$ for Venus could be 0.9 to 1; conversely, if abiogenesis on Earth was a rare event or if early Venus was more different from early Earth than current models constrain, $O_A$ may be 0.1 or less.

$O_P$ in our solar system may be nonzero from possible transportation of life due to impacts ejecting material from Earth, known to have occurred semi-regularly (Nicholson, 2009; Beech *et al.*, 2018). The potential for panspermia in the solar system has been investigated in some detail (von Hegner *et al.*, 2020a; b), and highlights the importance of assessing overlapping periods of habitability between worlds transporting viable life, as well as the ability of life to survive the transportation process (González-Toril *et al.*, 2005). The subfactor $O_P$ could thus be broken down into multiple subfactors of its own. For our purposes, it is sufficient to note that because of its relative proximity and size, Venus is the most likely potentially habitable body to receive viable life dislodged from Earth by large impacts (Gladman *et al.*, 1996), and thus, over geologic time is the most likely body in the solar system outside Earth to have an $O_P > 0$.

Just seeding viable life, however, would not be enough to result in a globally-distributed, sustained biosphere. For assessing the likelihood of life that we might be able to detect with astrobiological investigations, *breakout* ($O_B$) is an essential component of origination. It is the chance that life expanded beyond its point of origin to spread widely across the planet. On Earth, life may have arisen once, dozens or thousands of times in different surface or subsurface regions only to be snuffed out by events for which we have no record. This is encapsulated in the variable $O_B$, which for Earth, empirically, became 1 early in its history.

Accounting for breakout modifies the Origination term (Eq. 2):

$$O = (O_A + O_P - (O_A \bullet O_P)) \bullet O_B \qquad (3)$$

Following the same reasoning as with $O_A$ based on modelled similarities between early Earth and early Venus, the upper bound of $O_B$ for Venus could be 0.9 to 1, and the lower bound 0.1 or less as a starting probability span for life getting a foothold on the second planet.

*Planetary and Astrobiology Study of Origination*: Origination of life is a fundamental focus of evolutionary biology and astrobiology alike. Understanding the possibility of independent origination of life on Venus benefits substantially from investigations of life's origins on Earth, including how similar conditions on early-Venus were to Archaean and Hadean Earth. Determining the likelihood of transfer from Earth or other life abodes to Venus would benefit from statistical modeling of the transfer and survivability of life as we understand it between the solar system's rocky planets (e.g., Nicholson *et al.*, 2009). Favorable biochemical conditions on several ocean worlds in the outer solar system may have allowed life to arise, but those bodies would have a lower chance of exchanging materials with rocky planets like Earth and Venus due to their distance. Observation and modeling of the frequency of atmospheric impacts within solar systems (e.g., Harrington *et al.*, 2004), and further exploration of potential abodes and possible life cycles on Venus in particular (Seager *et al.*, 2016; 2020), will also help constrain the range of values of the Origination factor.

*Why O is not zero for Venus:* Current models suggest that early Venus conditions paralleled those of early Earth during the period in which Earth life arose (Way *et al.*, 2016; Way and Del Genio, 2020), which, absent other information, supports an Earth-like value of $O_A = 0.9 \sim 1$. Regardless of a potential independent biogenesis on Venus, we know that lithopanspermia subfactors outlined in $O_P$ should have sent endolithic terrestrial microbes towards Venus throughout its habitable history (Beech *et al.*, 2018), implying that $O_P$ could be greater than 0. Breakout, also assuming habitable conditions paralleling early Earth as our model, is also estimable at greater than zero. With at least one of $O_A$ and $O_P$ nonzero, and $O_B$ nonzero, it thus follows that $O > 0$ for Venus.

**Robustness:** Life on Earth arose relatively early in the planet's history, and persisting through catastrophic asteroid impacts, global glaciation events, the oxygenation of the atmosphere, and many other challenges severe enough to cause mass extinctions (Hoffman *et al.*, 1998, 2002; Melezhik, 2006). Life spread so widely, quickly, and with such variety and quantity that the resulting biosphere was robust enough to not be completely eradicated in the face of both acute and gradual environmental changes, due both to external forcings and internal planetary changes. An estimation of this robustness may be expressed with:

$$R = R_B \bullet R_D \qquad\qquad (4)$$

where $R_B$ is a measure of potential biomass supported over time, and $R_D$ is a similar measure of potential biodiversity supported over time. The lower the value of R, the smaller and more fragile the biosphere is to endure the losses or threats captured in the final term of the VLE, Continuity (see below).

At its most abstract, R can be considered to represent the "best case" for a planetary biosphere at a given time. On Earth, our only example of a biosphere, the value of R has been sufficiently high to allow survival through dramatic climate events, near-global mass extinctions, and other regional changes delivering stress or pressure on ecosystems. We can estimate the lower end of an Earth-like planet's R by extrapolating from studies of life in extreme environments, comparative ecology, paleoclimatology, etc., as reference points for less-favorable global conditions. However, it is much more difficult to make conservative estimates about a planet theoretically *more* habitable than Earth. To account for these limits, we define R here as a fraction of $R_{Earth}$, and let $R_{Earth} = 1$. This makes our bias explicit, and allows a straightforward recalculation if one wishes to make different assumptions about Earth's relative habitability.

Conservatively, $R_{Earth}$ should represent Earth's total biosphere. However, a less stringent approach might use the metrics of Earth's closest analogue environment as the divisor instead. This would be justified in cases where (1) the potentially habitable environments of the target planet are relatively uniform (as is the case for Venus's aerosol layer, at least compared to the Venus surface and subsurface conditions), and (2) the corresponding Earth analogue environment is well-isolated, such as a subglacial lake, or has persisted as a habitat over long periods of time without re-seeding from other habitats, such as Earth's ocean.

The first subfactor of R is $R_B$, a measure of the amount of life present on a planet. In keeping with our strategy of normalizing these terms to $R_{Earth}$, we will first discuss past and present estimates of Earth's biomass, and then move on to possible past and present Venus biomass.

Biomass on Earth, on a planetary scale, is usually quantified as organically bound carbon measured in gigatons (Gt) C. This definition assumes both a biosphere based on carbon and that the majority of organic carbon is biogenic. Earth's total biomass also includes large contributions from multicellular organisms, particularly plants. None of these are necessarily true for other potential biospheres (NRC, 2007). However, part of the astrobiological appeal of Venus is its similarity to Earth during the early period in which life may have arisen, implying a similar potential biochemistry. We therefore use estimates of Earth's microbial $R_B$ in Gt C as our divisor. In the modern era, bacteria and archaea on Earth are the second largest global biomass component (~77 Gt C), the vast majority of which reside in subsurface environments (Bar-On *et al.*, 2018). Although much remains unknown about Earth's Archaean biosphere, which was entirely microbial and aquatic, it has been suggested that the warmer oceans and higher $CO_2$ levels supported similar or even higher levels of microbial biomass (Franck *et al.*, 2005).

For early Venus, consistent with our approach for the O factor, let us assume that its planetary history was relatively similar to early Earth's, at least in the initial 1-2 Ga, and that a potentially similar biochemistry developed. The maximum size of an early Venus biosphere modelled on Earth's Archaean biosphere would then be constrained by the size of potential Venus aquatic habitats (oceans, lakes, groundwater) among other factors such as temperature, atmospheric composition, and levels of biologically important solutes such as nitrates and phosphates. These factors are not currently well constrained; for example, models of early Venus have used a water presence ranging from a global ocean covering ~60% of the surface -- not dissimilar to Earth's 70% -- to sparse groundwater -- less than 0.05% (Way & Del Genio, 2020). In the absence of information on other factors, we use early Earth as a template, yielding a range of values of $R_B$ for early Venus of 0.0005 – 0.85.

For the aerial biosphere hypothesis on modern-day Venus, we can place an order-of-magnitude upper bound on $R_B$ with a thought experiment that assumes that *all* the particles in Venus's cloud layer larger than 0.2 μm (the lower end of terrestrial microbes' size range) are microorganisms. A quick calculation using particle concentrations summarized from in-situ measurements (Esposito *et al.*, 1984) yields a count of $5 \times 10^{24}$ potential organisms. By comparison, the estimated number of prokaryotes on Earth currently is in the range of $4 \times 10^{30}$ (Whitman *et al.*, 1998). If we assume an optimistic range of $10 – 75$ fg C cell$^{-1}$ (Kallmeyer 2012, Cermack 2017), the present day Venus clouds yield a biomass range of $0.00005 – 0.0004$ Gt C, or an $R_B$ of $7 \times 10^{-7} – 5 \times 10^{-6}$. These assumptions yield a very low, but still nonzero, upper limit on the current $R_B$ of Venus. The small total volume of Venus's aerosols, as derived from the sizes and distributions measured to date, is a significant constraint for this scenario. The quantity, nature, and variation of these aerosols (Figure 3) could be further constrained by in-situ atmospheric observation.

The less conservative approach of comparing the potential habitat of Venus's clouds to Earth's closest analogue environment is complicated by our limited understanding of Earth's aerobiosphere. Airborne transit of microbes occurs at a global scale (Schuerger *et al.*, 2018), and prior sampling efforts indicate that the total number of viable cells in the atmosphere may be on the order of $5 \times 10^{20}$, with ~90% in fog and cloud water (Fuzzi *et al.*, 1997; Harris *et al.*, 2002; Amato *et al.*, 2005; Smith *et al.*, 2018; Bryan *et al.*, 2019). However, traces of metabolic activity have been observed so far only in warm, wet cloud droplets near the surface (Amato *et al.*, 2019), and reproduction while airborne has not yet been observed in-situ; at higher-altitude regions more isolated from surface sources, recovered cells are desiccated and dormant (Bryan *et al.*, 2019). If we assume that Earth does have a persistent, if low-level, aerobiosphere, then using its biomass as the divisor for the same upper bound on Venus biomass calculated above yields a value three orders of magnitude larger than 1. (This is expected, as an upper bound on Earth's atmospheric biomass based on Earth's water vapor volume would similarly exceed the actual estimated value). This alternate, non-conservative approach to $R_B$ would therefore give a final value of 1.

The second subfactor of R is $R_D$, a measure of the diversity of life present. In keeping with our strategy of normalizing these terms to $R_{Earth}$, we will first discuss past and present estimates of Earth's biodiversity, and then move on to possible past and present Venus biodiversity.

Life on Earth is incredibly diverse, with nearly every liquid and solid surface colonized with a detectable microbial population. Taxonomic classification of bacteria and archaea is a rapidly evolving field, and by some estimates 99.99% of microbes in most field samples belong to unknown taxa regardless of which diversity index is used to catalog species; estimates of the number of microbial species distributed across Earth vary, with conservative estimates on the order of $10^{12}$ (e.g., Locey and Lennon, 2016). This estimate does not include the number of microbial species that may have arisen and gone extinct over Earth's history, and we know even less about the biodiversity of Archaean Earth.

Quantitative metrics of taxonomic diversity, such as abundance measures, are therefore unlikely to be helpful in modeling other theoretical biospheres. Instead, functional diversity, which reflects how many distinct niches life occupies in a given habitat -- e.g., "apex predator (obligate heterotroph)" or "primary producer (sulfur-reducing chemolithotroph)" -- is probably the most intuitively applicable to a theoretical, non-terrestrial biosphere. As with biomass, the metric chosen should reflect constrainable similarities between Earth and Venus. For hundreds of millions of years, early Earth lacked several major functional niches present today (Nisbet, 1995), including oxygenic photosynthesis and all, or nearly all, land-based ecology. Conversely, we currently have no way of knowing how many historical niches may once have existed -- e.g., chemolithotrophs utilizing minerals only stable in a reducing atmosphere -- but are now lost.

We know even less, of course, about early Venus. However, since R is meant to represent a best-case, and we know that any extant life on Venus is almost certainly a relic of a more thriving era, we might look to an early Earth-like range for $R_D$ and use this as our estimate for early Venus as well. The same caveats regarding factors needed to better understand $R_B$ on Venus apply.

Modern-day Venus is a more tractable case for $R_D$. Although Earth has no direct analogue environment resembling Venus's clouds (Figure 1), we can estimate an upper bound based on partial analogues. Chemically speaking, terrestrial acid hot springs have been proposed as the inhabited environments that most closely reproduce Venus cloud temperature (97 to -45 ºC) and pH (less than -1.3 to 0.35; Grinspoon and Bullock, 2007; Krasnopolsky, 2019). Terrestrial deserts or concentrated brines may best represent the low water activity in Venus aerosols (~0.02 at a relatively optimistic assumption of 75% $H_2SO_4$ and 25% $H_2O$; Deno and Taft, 1954; Hansen and Hovenier, 1974; Kieft, 2003; Bolhuis *et al.*, 2006). Each of these environments show significantly less diversity than more typical mesophilic environments. At pH levels at or below 1, terrestrial life is limited to a few lineages of archaea (Barrie Johnson and Hallberg, 2008). Brines at water activities below 0.75 are similarly limited to other lineages of archaea (Grant, 2004; Oren, 2011). Life forms in both environments rely upon "narrow" metabolic pathways (Barrie Johnson and Hallberg, 2008; Oren, 2011), that likely emerged via horizontal gene transfer with bacteria (Fütterer *et al.*, 2004; Sorokin *et al.*, 2017); and although these modern archaea are generally aerobic heterotrophs (not useful for analogies to Venus), Earth has been oxygenated for a sufficiently long time to make aerobic metabolisms evolutionarily favorable. Only the most extreme deserts on Earth approach water activities below 0.1, and here, though more taxonomically diverse, life is primarily phototrophic, adapted to long periods of inactive desiccation followed by brief spurts of activity during sporadic water influx. To give a sense of the possible range of values, the smallest possible non-zero value of $R_D$ is equivalent to the apparently natural monoculture found in a terrestrial deep subsurface fracture (Lin *et al.*, 2006); its long-term isolation from other habitats and energy/nutrient limitations have some similarities to the suggested relict Venus ecosystem. At the high end of the range, it has been suggested that Earth's total airborne biodiversity might be equivalent to that of soil (Brodie *et al.*, 2007), approaching an $R_D$ of 1; this is possible due to the long tail of rare species, despite the much lower biomass. The latter value is less representative of a Venus scenario, as it includes a large contribution from short-term bioaerosols kicked up from the well-populated surface and quickly deposited again. We would therefore expect $R_D$ for a Venus aerobiosphere to be significantly lower than $R_D$ for Earth, but still nonzero.

*Planetary and Astrobiology Study of Robustness:* The similarity between Venus and Earth's early history, and the drastic divergence between their current states, is significant in the study of rocky planets. Many of the questions surrounding the potential characteristics of an early Venus biosphere are the same as those we still seek to answer for early Earth: nutrient and energy sources and cycling, ocean breadth and depth, radiation flux, transport of atmospheric gases and particulates. The global imaging provided by Magellan show us that regions of tessera terrain are stratigraphically older than the bulk of the Venus surface and represent an extinct tectonic and possible mineralogic and/or weathering regime (e.g., Gilmore *et al.*, 2017). Additional imaging, spectroscopy, mineralogy and chemistry of the tesserae, in-situ measurement of noble gases to assess volatile inventory and assessment of current and past geologic activity are all necessary steps to constrain the volatile history of ancient Venus.

The case of modern-day Venus is also complicated by Earth's lack of a habitat directly analogous to Venus's cloud layer. The closest regime, in terms of chemistry and isolation from surface nutrient and water sources, is probably the stratospheric sulfate layer (Gentry and Dahlgren, 2019), where the longest-enduring microbe-sized aerosols may have residence times of many months. However, under stratospheric conditions, significantly less dense than the "gas ocean" of Venus, it is likely impossible for terrestrial microorganisms to metabolize, grow, or reproduce. Survivors recovered at extreme heights above the Earth's surface tend to be dormant, resilient, endospore-forming bacteria enduring harsh irradiation until dropping out by gravitational settling (Bryan *et al.*, 2019). At lower altitudes, terrestrial aerobiologists are exploring whether short-lived airborne ecosystems exist within Earth clouds where environmental conditions are more favorable, including water and nutrient availability (Amato *et al.*, 2019).

A better understanding of Earth's aerobiosphere, and whether it has been as consistent as Earth's surface or subsurface biospheres, might lead to an alternative value for $R_{Earth}$. More generally, further exploration and characterization of the biomass limitations and functional diversity of those partial Venus analogues that are more accessible on Earth will continue to help inform our estimates of $R_B$ and $R_D$ for both Earth and Venus.

The VLE is inherently a whole planet question, based on what we know, can know, and must surmise about the planet as an integrated whole. Factors like R, and subfactors like $R_D$ and $R_B$ can be further subdivided into smaller and smaller components, down to individual biomes or niches (Figure 4). While doing this many times across the Earth would, in aggregate, improve an understanding and estimate of global R for Earth, and concomitantly Venus, delving deeply into any one example at this high level can in fact be counterproductive to a global estimation. Thus, detailed analysis and interrelationships of individual niches (on Earth and potentially on Venus) can be part of the greater solution, but is beyond the scope of this overview.

*Why R Is Not Zero for Venus:* Because *R* represents a "best-case" biosphere, it could only be zero for a target environment which meets no known criteria for habitability - for example, the sun, dry lunar regolith, or the exposed surface of an asteroid. While the potentially supportable biosphere on modern Venus may be quite low or limited by terrestrial standards, the relative clemency of early Venus, and its similarity to the empirically inhabited early Earth, allows R to be greater than zero.

**Continuity:** This factor reflects the necessity of continuous existence of habitats over time and space; or, equivalently, the lack of global extinction-level events. This might naturally lead to subfactors for C of $C_T$ and $C_S$, for temporal and spatial continuity respectively, and further subdivision to individual biosignatures or disequilibrium chemistry (e.g., Wogan and Catling, 2020) but our current state of knowledge of Venus is not sufficient to allow us to mathematically relate them with confidence (conveyed in Figure 3). For life to exist on Venus into the present, both $C_T$ and $C_S$ must be nonzero - that is, from the Origin point for life, there must be an unbroken continuity of habitable condition in both time and space.

Environmental continuity is affected by both internal and external factors. The former includes variations in the carbonate/water/sulfur cycles that are governed by plate recycling, rates of volcanism and rock weathering. Study of the composition and deformation histories of the most ancient terrains on Venus (tesserae) may help determine the presence, extent and duration of some of these factors (e.g., Gilmore *et al.*, 2017). External factors include solar-system wide events such as stellar variability (life-threatening flares/coronal mass ejections), stellar aging (changing luminosity inducing climatic shifts), and large impactors (Bostrom and Circovic, 2011; Chapman and Morrison, 2013). Some of these external factors, such as stellar lifetime, are empirically known to be 1 for Venus by the continuing presence of life on neighboring Earth. Others, such as activity sufficient to sterilize life only out to 0.8 AU, or an extinction-level coronal mass ejection while Venus and Earth were in opposition, must be estimated.

Continuity can be quantifiably constrained for Venus through direct measurement; determining the availability of current resources in potential niches (e.g., the elements C, H, N, O, P, & S and solvents as necessary building blocks for Earth-like biology in Venus clouds), and through unraveling the geologic history of the planet to determine if a continuous path might have been available for life to evolve to survive and maintain itself for tens or hundreds of millions of years of post-ocean Venus history. For example, one possible pathway to extant Venus life would require conditions to evolve contiguously and continuously from a marine-land interface (e.g., one of the likely 'breakout' environments for Earth), to a globe-spanning biosphere, to eventual adaptation towards complete airborne life cycles and an aerobiosphere maintainable solely in the clouds. The identification and confirmation of potential biosignatures in the atmosphere of a planet (e.g., Kaltenegger, 2017; Wogan and Catling, 2020), would argue for a larger value for continuity. The recent discovery of phosphine in the Venus atmosphere (Greaves *et al.*, 2020), if verified as a biologically produced gas, would set C =1 for Venus.

However, complications for continuity are rooted in both understanding of terrestrial biology and lack of understanding of Venus' geologic history and present conditions. If one assumes a terrestrial-like biochemistry, neither the trace composition of Venus's aerosols nor current conditions such as atmospheric circulation of dust are understood well enough to determine water activity or the presence of bioavailable forms of nitrogen and phosphorus, let alone the enzymatically important heavy elements like Fe, Zn, Pb, Cu, Sn, V, Cd, Ni, Se, Mn, Co, Cr, As, Mo and W that must be available at low but consistent levels to terrestrial life. On Earth these are quite scarce in the atmosphere. However, in situ detection of both phosphorus and iron were reported by the Vega X-ray fluorescence spectrometers (Andreychikov *et al.*, 1987), and a comprehensive trace elemental assay of the Venus clouds, with sensitive 21$^{st}$ century instruments, has not been performed. Non-chemical requirements are also critical. Energy pathways such as photosynthesis or chemosynthesis need to have been established and maintained or evolved to in a similar contiguous and continuous manner. For example, although Venus receives more photonic energy at the top of its atmosphere than Earth, the thick atmosphere and haze layer reflect or block larger fractions of it (Titov *et al.*, 2007) thus the zonally averaged incoming flux at the Venus cloud tops is ~1.5× less than that for Earth. This is particularly important for a potential atmospheric ecosystem, as the residence time of potential aerosol habitats imposes a particular constraint (Seager *et al.*, 2020). Many terrestrial microbes in extremely harsh or nutrient-limited environments have very long generation times of weeks to months, potentially in combination with long periods of inactivity. Seager *et al.* (2020) suggest that updrafts and/ or gravity waves on Venus may be sufficient to return desiccated life in the haze layer to conditions where they can thrive and continue their life cycle.

One of the major sub-factors of C specific to Venus (affecting continuity in both space and time) is the timeline of Venus's water loss and cloud formation. Although life is capable of very rapid adaptation and diversification in some circumstances, major habitat transitions on Earth such as colonization of land took at least hundreds of millions of years. The shorter the periods of overlap between origination of life in the oceans, the evaporation of the oceans into a water-cloud-shrouded planet, and the formation of the modern-day habitat of persistent sulfuric cloud cover, the lower the likelihood of colonization; and if any two did not overlap at all, it might be negligible at best. What matters to estimating C is knowing the history of the planet; its geological and climate evolution. For example, Venus' water history timeline is not currently well constrained, and some models include the possibility of a gap of 10s Myr to 100s Myr between the end of Venus' surface oceans and the current state of a persistent cloud deck of sulfuric acid aerosols (Bullock and Grinspoon, 2001). The duration of that gap (if it was present), or its absence, could significantly affect the estimation of C.

*Planetary and Astrobiology Study of Continuity*: Continuity is hard to estimate given how little we know about both Venus' history and current potential habitats and the potential biases the N=1 of Earth engenders. For example, what is observed to be an "essential" element in the terrestrial biota is also the product of an opportunistic evolutionary process which might well have found "work arounds" in other planetary environments with a different complement of available elements. More exotic proposed biochemistries (such as direct use of sulfuric acid as an alternative polar solvent) are even less constrained in terms of energy and chemistry requirements (Schulze-Makuch and Irwin, 2008; Cockell and Nixon, 2016).

However, of the three factors of the VLE, Continuity is the one we can do the most to improve quantification through direct study of Venus. Most of the areas delineated in the Goals, Objectives, and Investigations for Venus Exploration (VEXAG, 2019) document will result in direct quantitative improvement of the estimate for C. For example, determining the presence and extent of silicic igneous rocks constrains the history of possible early Venus oceans and crustal evolution. Measuring isotopic ratios of noble gases, oxygen, hydrogen in the atmosphere will constrain the history of water, and possibly biological or prebiotic effects on global chemistry. If microbe-bearing aerosols, on average, settle out (as on Earth) or fall to an altitude at which they dry out or boil off (as on Venus) faster than the microbes can reproduce, an aerosol-based biosphere without periodic injections from other reservoir habitats will not be stable over the long term, even if short-term conditions are otherwise favorable. Deep dynamics will constrain the possibility of circulation of materials from near the surface through the lower atmosphere, and geologic history and activity will determine the present and past supply of chemicals to different parts of potential Venus ecosystems. The entirety of VEXAG Goal 1, in fact, prioritizes the understanding of Venus' early history and potential habitability. And certainly, verification of the presence and a biogenic source for phosphine (Greaves *et al.*, 2020) would result in C=1.

*Why C Is Not Zero for Venus.* At the moment, this could be the most difficult stipulation of the VLE. We simply don't know enough about Venus' evolution to do more than make model and geologically and evolutionary plausible 'what if' scenarios. It is, however, the most amenable factor to quantitative improvement. Continuity estimates can only be vastly improved with each new mission we send to Venus.

**Life:** The arguments presented for each of the factors and major subfactors of the Venus Life Equation (VLE) being nonzero encourage varying estimations for L. We can make illustrative examples of "low and "high" calculations. The numbers presented here are not meant to be taken as predictions, and are used to show the potential range of consideration and identify some of the major constraints worth further study. Thorough examination of existing literature may lend to more conservative or generous predictions of one or more factors. Additional research of Earth and Venus systems in-situ and in the laboratory will all help refine and improve estimates of one or more factors.

Our original, motivating hypothesis is the likelihood of an extant, airborne, microbial biosphere in Venus's cloud layers. Combining the above estimates for all 3 factors, we can calculate a range for the chance of life existing today. For Origination, a "high" chance of life getting started on Venus might be 100% as some combination of the chance of abiogenesis and/or panspermia - if one or the other is considered "more likely than not" and breakout is thought of in similar ways as breakout on Earth. A "low" O term might be 0.1, for converse reasons. Similarly, thinking of Robustness in a "best case" aerobiosphere allows an R range of $7\times10^{-7} - 0.1$; based upon the number of particles present in modern day Venus and different assumptions about how similar were ancient Venus and Earth. Given the knowledge gaps of Venus' geologic and climate history, assigning for C a range of 0.1 to 1.0 are currently appropriate. Using example low and high values throughout:

$$L = 0.1 \bullet 7\times10^{-7} \bullet 0.1 = 7\times10^{-9} \text{ (low) or} \qquad (7)$$

$$L = 1.0 \bullet 0.1 \bullet 1.0 = 0.1 \text{ (high)} \qquad\qquad (8)$$

As a demonstration of the flexibility of the Venus Life Equation, we can repeat the same calculation for the question of whether life *ever* existed on Venus. Estimates of Origination are the same in both cases. Robustness is at a maximum for early Venus, using Archaean Earth as a guide, estimated as 0.2. C will be left as a range of 0.1 to 1.0, recognizing that early Venus may have been quite Earth-like. Using example low and high values throughout:

$$L = 0.1 \bullet 1\times10^{-4} \bullet 0.1 = 1\times10^{-6} \text{ (low) or} \qquad\qquad (9)$$

$$L = 1.0 \bullet 1.0 \bullet 1.0 = 1.0 \text{ (high)} \qquad\qquad (10)$$

These numbers are simply example calculations that can and should be refined by others.

| Parameter | | Modern Venus | | Ancient Venus | |
|---|---|---|---|---|---|
| | | Estimate | Rationale | Estimate | Rationale |
| **O** | **Low** | 0.1 | Not Earthlike, or not earthlike for long, or abiogenesis is rare | 0.1 | Not Earthlike, or not earthlike for long, or abiogenesis is rare |
| | **High** | 1 | Long habitable environment, origination events are common | 1 | Long habitable environment, origination events are common |
| **R** | **Low** | $10^{-7}$ | Earth's microbial biomass vs. maximum possible Venus "aerosol" biomass; Venus much lower diversity than Earth | $10^{-4}$ | Venus not as hospitable as Archaean Earth, low water scenario |
| | **High** | 0.1 | Earth's airborne biomass vs. maximum possible Venus "aerosol" biomass; Venus lower diversity than Earth | 1 | Long habitable environment with abundant niches |
| **C** | **Low** | 0.1 | Significant continuity gaps | 0.1 | Significant continuity gaps |
| | **High** | 1 | No continuity gaps (like Earth) | 1 | No continuity gaps (like Earth) |

Individual investigations and additional constraints, experiments, or observations may drive any one of the factors or subfactors higher, or perhaps more likely, lower. This exercise can be performed for any potential abode of life in our solar system and adapted and estimated for any potential habitable world. For example, the known and theorized sub-ice oceans on several icy moons (and Pluto) harbor several niches and pathways that might increase R or C in the equation, with $O_p$ from the inner solar system being statistically smaller. The atmospheres of the giant planets require a different set of assumptions for sources and renewability of potential heavier elements for life processes. In all these cases, for our own solar system, we know where to look and what we can measure to more quantitatively constrain these factors.

Although a similar exercise may well reveal several environments on solar system bodies having ranges of estimated L which exceed that of Venus, any strategy for astrobiology exploration must also factor in the accessibility of the potentially habitable environment. Venus is the nearest planet to Earth in both average distance and delta-V. A non-negligible estimate for L on Venus would seem to argue strongly for exploratory missions operating within the cloud environment as part of any comprehensive strategy to look for extant life in the solar system.

It is worth noting that estimates of both R and C suffer from observability bias. Though we can speculate about potential habitats by extrapolating from Earth life, the only thing that ultimately proves habitability is inhabitation. If any R or C term could be directly measured, we would automatically have an empirical value of one or zero for L. The VLE is intended to be an exercise in identifying assumptions and needed constraints for a planet for which direct life detection efforts have yet to be attempted.

*Beyond Venus:* The L determined by the Venus Life Equation, adapted for and integrated over many possible worlds, is related to the term $f_l$ of the Drake Equation (Burchel, 2006): the fraction of planets in our galaxy that develop life. The equation applied to Venus shows how we might approach questions of habitability on worlds beyond Earth. Further exploration of our own system's potential abodes, and examination of the properties and statistics of the growing number of known planets outside the solar system (Seager *et al*., 2016; Rossmo 2017); may allow an expansion from consideration of Venus only toward a "Planetary Life Equation" to better estimate the probability of life on exoplanets (Catling *et al.*, 2018).

*Consequences for Planetary Protection:* Currently, NASA classifies Venus missions under planetary protection Category II, which "includes all types of missions to target those bodies where there is significant interest relative to the process of chemical evolution and the origin of life, but where there is only a remote chance that contamination carried by a spacecraft could jeopardize future exploration," (NAS 2006). The National Academies (2006) recommended that the Category II planetary protection classification of Venus be retained. With respect to forward contamination of Venus clouds, this recommendation is based on the conclusion that "the cloud droplets consist of concentrated sulfuric acid, any terrestrial organisms would be rapidly destroyed by chemical degradation."

The only terrestrial life that might endure conditions in Venus aerosols are extreme acidophiles, which have not been observed to survive long periods of time airborne and are unlikely to be spacecraft assembly facility contaminants (Smith *et al.*, 2017). This is true even if one assumes that putative extant Venus microbes rely on substantially different metabolic inputs and outputs from possible transported terrestrial life to the point that the "potentially habitable region" for each does not overlap. This implies that the types of possible terrestrial (bio)chemical contamination that could survive exposure to Venus atmosphere/aerosols are unlikely to cause false positives in experiments looking for Venus life. The NAS study did not recommend any scientific investigations for the specific purpose of reducing uncertainty with respect to planetary protection issues. Thus, the Category II classification of the NAS study remains as yet unchallenged by a nonzero value for L. Like any life detection experiment, however, any in-situ instrumentation will need to invoke a high level of sterilization and cleanliness to ensure accurate measurement.

**Conclusion:** The Venus Life Equation allows an estimation of the likelihood of extant life on Venus can be made by examining the Earth analog, improving our understanding of planetary bodies in our solar system, and studying current conditions on Venus. The chances of life originating and surviving on Venus to today are low, but nonzero. Improved in-situ observation of conditions on Venus, especially in the potentially habitable zone in its middle atmosphere will help to constrain any estimate for Venus and other cloudy worlds. Adaptation and expansion of the Origination, Robustness, and Continuity factors to additional bodies within and outside our solar system will allow adaptation of the equation's principles to other potential biospheres.

**Acknowledgements**: This work benefited from discussions across several years within NASA's Venus Exploration Analysis Group (VEXAG), the Venera Landing Site and Cloud Habitability Workshop (October, 2019, Moscow, Russia) and the Exoplanets in our Backyard workshop (February 2020, Houston, TX). We appreciate two careful reviews that helped improve this manuscript.

**References:**

Amato P, Ménager M, Sancelme M, Laj P, Mailhot G, Delort AM. Microbial population in cloud water at the Puy de Dôme: implications for the chemistry of clouds. *Atmospheric Environment*. 2005 Jul 1;39(22):4143-53.

Amato P, Besaury L, Joly M, Penaud B, Deguillaume L, Delort AM. Metatranscriptomic exploration of microbial functioning in clouds. *Scientific Reports 2019*;9(1):4383. doi.org/10.1038/s41598-019-41032-4

Andreychikov BM, Akhmetshin IK, Korchuganov BN, Mukhin LM, Ogorodnikov BI, Petryanov IV, Skitovich VI. X-ray radiometric analysis of the cloud aerosol of Venus by the Vega 1 and 2 probes. *Cosmic Res* 1987;25:16.

Arking A, Absorption of solar energy in the atmosphere: Discrepancy between model and observations. *Science* 1996; 273(5276);779-782.

Bar-On YM, Phillips R, Milo R. The biomass distribution on Earth. *P Natl Acad Sci* 2018;115 (25):6506–11. https://doi.org/10.1073/pnas.1711842115.

Barrie Johnson D, Hallberg KB. Carbon, iron and sulfur metabolism in acidophilic micro-organisms. *Adv Microb Physiol* 2008;54:201–55. doi.org/10.1016/S0065-2911(08)00003-9

Beech M, Coulson IM, Comte M. Lithopanspermia – The terrestrial input during the past 550 million years. *Am J Astron Astroph* 2018;6(3): 81-90.

Bolhuis H, Palm P, Wende A, Falb M, Rampp M, Rodriguez-Valera F, Pfeiffer F, Oesterhelt D. The genome of the square archaeon Haloquadratum walsbyi: Life at the limits of water activity. *BMC Genomics* 2006;7(1),1. doi.org/10.1186/1471-2164-7-169

Bostrom N, Cirkovic MM, eds. *Global catastrophic risks*. 2011 Oxford University Press.

Brodie EL, DeSantis TZ, Parker JP, Zubietta IX, Piceno YM, Andersen GL. Urban aerosols harbor diverse and dynamic bacterial populations. *Proceedings of the National Academy of Sciences*. 2007 Jan 2;104(1):299-304.

Bryan NC, Christner BC, Guzik TG, Granger DJ, Stewart MF. Abundance and survival of microbial aerosols in the troposphere and stratosphere. *ISME Journal* 2018;13,11:2789-2799.

Budisa N, Schulze-Makuch D. Supercritical carbon dioxide and its potential as a life-sustaining solvent in a planetary environment." *Life* 2014;4 (3):331–40. doi.org/10.3390/life4030331

Bullock MA, Grinspoon DH. The recent evolution of climate on Venus. *Icarus*. 2001 Mar 1;150(1):19-37.

Burchell MJ. W (h) ither the Drake equation? *Int J Astrobio* 2006;5(3):243-250.

Catling DC, Krissansen-Totton J, Kiang NY, Crisp D, Robinson TD, DasSarma S, Rushby AJ, Del Genio A, Bains W, Domagal-Goldman S. Exoplanet biosignatures: a framework for their assessment. *Astrobiology*. 2018 Jun 1;18(6):709-38.

Cavicchioli R. Extremophiles and the search for extraterrestrial life. *Astrobiology* 2002;2,3:281-292.

Cermak, Nathan, Jamie W. Becker, Scott M. Knudsen, Sallie W. Chisholm, Scott R. Manalis, and Martin F. Polz. "Direct Single-Cell Biomass Estimates for Marine Bacteria via Archimedes' Principle." The ISME Journal 11, no. 3 (March 2017): 825–28. https://doi.org/10.1038/ismej.2016.161

Chapman CR, Morrison D, eds. *Cosmic catastrophes*. 2013 Springer.

Cockell CS, 1999. Life on Venus. *Planet Space Sci* 1999;47(12):1487–1501. doi.org/10.1016/S0032-0633(99)00036-7.

Damer B, Deamer D, The Hot Spring Hypothesis for an Origin of Life. Astrobiology 2020; 20 (4): 429-452.doi: 10.1089/ast.2019.2045

Deno NC, Taft RW. Concentrated sulfuric acid-water. *J Am Chem Soc* 1954;76(1):244–248. doi.org/10.1021/ja01630a063

Drake FD, The Radio Search for Intelligent Extraterrestrial Life, G. Mamikunian and M.H. Briggs (Eds.), *Current Aspects of Exobiology*, Pergamon Press,1965, pp. 323-345.

Esposito LW, Knollenberg RG, Marov MYa, Toon OB, Turco RP. The clouds and hazes of Venus. *Venus*, Univ. Ariz. Press, Humten DM, Colin L, Donahue TM, Moroz VI. eds. 1984;16:484-564.

Franck S, Bounama C, von Bloh W. Causes and timing of future biosphere extinction. *Biogeosci Disc* 2005;2(6): 1665–79. doi.org/10.5194/bgd-2-1665-2005.

Fütterer O, Angelov A, Liesegang H, Gottschalk G, Schleper C, Schepers B, Dock C, Antranikian G, Liebl W. Genome sequence of picrophilus torridus and its implications for life around pH 0. *P Natl Acad Sci* 2004;101(24):9091–96. doi.org/10.1073/pnas.0401356101.

Fuzzi S, Mandrioli P, Perfetto A. Fog droplets—an atmospheric source of secondary biological aerosol particles. *Atmospheric environment.* 1997 Jan 1;31(2):287-90.

Jacobson MZ. Development and application of a new air pollution modeling system—Part III. Aerosol-phase simulations. Atmospheric Environment. 1997 Feb 1;31(4):587-608.Gánti T, *The principles of life*, 2003, Oxford University Press.

Gentry D, Dahlgren RP. Venus aerosol sampling considerations for in situ biological analysis. *The Venera-D Landing Sites and Cloud Habitability Workshop* 2019: Moscow, Russia.

Gilmore MS, Treiman A, Helbert J, Smrekar S, Venus surface composition constrained by observation and experiment. *Space Sci. Rev.,* 2017;11:(1-30). doi:10.1007/s11214-017-0370-8.

Gladman BJ, Burns JA, Duncan M, Lee P, Levinson HF, The exchange of impact ejecta between terrestrial planets. *Science* 1996;271:1387–1392.

González-Toril E, Martínez-Frías J, Gómez Gómez JM, Rull F, Amils R, Iron meteorites can support the growth of acidophilic chemolithoautotrophic microorganisms. *Astrobiology*, 2005, 5(3), pp.406-414.

Grant WD. Life at low water activity. *Phil Trans Royal Soc Bio Sci* 2004;359(1448):1249–67. doi.org/10.1098/rstb.2004.1502.

Greaves GS, Richards AMS, Bains W, Rimmer PB, Sagawa H, Clements DL, Seager S, Petkowski JJ, Sousa-Silva C, Ranjan S, Drabek-Maunder E, Fraser HJ, Cartwright A, Mueller-Wodarg I, Zhan Z, Friberg P, Coulson I, Lee E, Hoge J, Phosphine gas in the cloud decks of Venus, *Nature Astronomy*, 2020, pp. 1-10

Grinspoon DH, Bullock MA. Astrobiology and Venus exploration." *Geophys Monograph Series* Esposito LW, Stofan ER, Cravens TE eds. 2007;176:191–206. Washington, D. C.: AGU. doi.org/10.1029/176GM12.

Hansen JE, Hovenier JW. Interpretation of the polarization of Venus. *J Atmo Sci* 1974;31:1137–1160. doi.org/10.1175/1520-0469(1974)031

Harrington JdPI, Brecht SH, Deming D, Meadows V, Zahnle K, Nicholson PD. Lessons from Shoemaker-Levy 9 about Jupiter and planetary impacts. *Jupiter: The Planet, Satellites and Magnetosphere.* F. Bagenol, T. Dowling, W. McKinnon, eds. Cambridge Univ. Press. 2004;8:158-84.

Harris MJ, Wickramasinghe NC, Lloyd D, Narlikar JV, Rajaratnam P, Turner MP, Al-Mufti S, Wallis MK, Ramadurai S, Hoyle F. Detection of living cells in stratospheric samples.

InInstruments, Methods, and Missions for Astrobiology IV 2002 Feb 5 (Vol. 4495, pp. 192-198). International Society for Optics and Photonics.

Hoffman PF, Kaufman AJ, Halverson GP, Schrag DP. A Neoproterozoic snowball earth. science. 1998 Aug 28;281(5381):1342-6.

Hoffman, P.F. and Schrag, D.P., 2002. The snowball Earth hypothesis: testing the limits of global change. *Terra nova*, *14*(3), pp.129-155.

Kallmeyer, J., R. Pockalny, R. R. Adhikari, D. C. Smith, and S. D'Hondt. "Global Distribution of Microbial Abundance and Biomass in Subseafloor Sediment." Proceedings of the National Academy of Sciences 109, no. 40 (October 2, 2012): 16213–16. https://doi.org/10.1073/pnas.1203849109.

Kaltenegger, L. (2017). How to characterize habitable worlds and signs of life. *Annual Review of Astronomy and Astrophysics*, *55*, 433-485.

Kieft TL. Desert environments: Soil microbial communities in hot deserts. Bitton G ed. *Encyclopedia of Environmental Microbiolog*y Wiley 2003;1–25. doi.wiley.com/10.1002/0471263397.env178

Krasnopolsky VA. *Spectroscopy and Photochemistry of Planetary Atmospheres and Ionospheres*. Cambridge University Press 2019;243. doi. 10.1017/9781316535561

Limaye SS, Mogul R, Smith DJ, Ansari AH, Słowik GP, Vaishampayan P. Venus' spectral signatures and the potential for life in the clouds. *Astrobiology* 2018;18(9):1181–98. doi.org/10.1089/ast.2017.1783

Lin LH, Wang PL, Rumble D, Lippmann-Pipke J, Boice E, Pratt LM, Lollar BS, Brodie EL, Hazen TC, Andersen GL, DeSantis TZ. Long-term sustainability of a high-energy, low-diversity crustal biome. *Science*. 2006 Oct 20;314(5798):479-82.

Locey KJ, Lennon JT, Scaling laws predict global microbial diversity. *PNAS* May 24, 2016;113 (21) 5970-5975 doi.org/10.1073/pnas.1521291113

Melezhik VA. Multiple causes of Earth's earliest global glaciation. Terra Nova. 2006 Apr;18(2):130-7.

Morowitz H, Sagan C. Life in the clouds of Venus. *Nature* 1967;215:1259-1260.

NAS (National Academies of Sciences Engineering and Medicine). Assessment of Planetary Protection Requirements for Venus Missions: Letter Report (2006). *National Academies Press*, 2006;16p.

Nicholson WL. Ancient micronauts: interplanetary transport of microbes by cosmic impacts. *Trends Microbio* 2009;17(6):243-350.

Nisbet EG. Archaean ecology: A review of evidence for the early development of bacterial biomes, and speculations on the development of a global-scale biosphere. *Geological Society, London, Special Publications* 1995;95(1):27–51. doi.org/10.1144/GSL.SP.1995.095.01.03

NRC (National Research Council). The limits of organic life in planetary systems. Washington, D.C.: *National Academies Press* 2007. doi.org/10.17226/11919.

Oren A. Ecology of halophiles. *Extremophiles Handbook* 2011;1:344–61. Springer. doi.org/10.1007/978-4-431-53898-1_3.4

Rossmo DK. Bernoulli, Darwin, and Sagan: the probability of life on other planets. *Intl J Astrobio* 2017; 16(2):185-9.

Schuerger AC, Smith DJ, Griffin DW, Jaffe DA, Wawrik B, Burrows SM, Christner BC, Gonzalez-Martin C, Lipp EK, Schmale III DG, Yu H. Science questions and knowledge gaps to study microbial transport and survival in Asian and African dust plumes reaching North America. Aerobiologia. 2018 Dec 1;34(4):425-35.

Schuerger AC, Smith DJ, Griffin DW, Jaffe DA, Wawrik B, Burrows SM, Christner BC, Gonzalez-Martin C, Lipp EK, Schmale III DG, Yu H. Science questions and knowledge gaps to study microbial transport and survival in Asian and African dust plumes reaching North America. Aerobiologia. 2018 Dec 1;34(4):425-35.Schulze-Makuch D, Grinspoon DH, Abbas O, Irwin LN, Bullock MA. A sulfur-based survival strategy for putative phototrophic life in the Venusian atmosphere. *Astrobiology* 2004;4(1): 11–18. doi.org/10.1089/153110704773600203.

Schulze-Makuch D, Dohm JM, Fairén AG, Baker VR, Fink W, Strom RG. Venus, Mars, and the ices on mercury and the moon: Astrobiological implications and proposed mission designs. *Astrobiology* 2005;5(6):778–95. doi.org/10.1089/ast.2005.5.778.

Schulze-Makuch D, Irwin LN. Definition of life. In *Life in the Universe* 2008;7-24. Springer, Berlin, Heidelberg.

Seager S, Bains W, Petkowski JJ. Toward a list of molecules as potential biosignature gases for the search for life on exoplanets and applications to terrestrial biochemistry. *Astrobiology* 2016;16(6):465-85.

Seager S, Petkowski JJ, Gao P, Bains W, Bryan NC, Ranjan S, Greaves J. The Venusian lower atmosphere haze as a depot for desiccated microbial life: a proposed life cycle for persistence of the Venusian aerial biosphere", *Astrobiology,* 2020; 18 p. https://doi.org/10.1089/ast.2020.2244

Smith DJ. Microbes in the upper atmosphere and unique opportunities for astrobiology research. *Astrobiology* 2013;13(10): 981-990.

Smith DJ, Ravichandar JD, Jain S, Griffin DW, Yu H, Tan Q, Thissen J, Lusby T, Nicoll P, Shedler S, Martinez P. Airborne bacteria in earth's lower stratosphere resemble taxa detected in the troposphere: Results from a new NASA Aircraft Bioaerosol Collector (ABC). *Frontiers in Microbiology*. 2018 Aug 14;9:1752.

Smith SA, Benardini III JN, Anderl D, Ford M, Wear E, Schrader M, Schubert W, DeVeaux L, Paszczynski A, Childers SE. Identification and characterization of early mission phase microorganisms residing on the Mars Science Laboratory and assessment of their potential to survive Mars-like conditions. *Astrobiology*. 2017 Mar 1;17(3):253-65.

Sorokin DY, Messina E, Smedile F, Roman P, Damsté JSS, Ciordia S, Mena MC, *et al.* Discovery of anaerobic Lithoheterotrophic Haloarchaea, ubiquitous in hypersaline habitats. *The ISME Journal* 2017;11(5):1245–60. doi.org/10.1038/ismej.2016.203.

Titov DV, Bullock MA, Crisp D, Renno NO, Taylor FW, Zasova LV 2007. Radiation in the atmosphere of Venus. *Geophys. Mono. AGU*, 2007, 176, p.121.

VEXAG (Venus Exploration Analysis Group). Venus Goals, Objectives, and Investigations 2019 https://www.lpi.usra.edu/vexag/reports/VEXAG_Venus_GOI_Current.pdf

von Hegner I, Extremophiles: a special or general case in the search for extra-terrestrial life? *Extremophiles*, 2020a; 24: 167-175.10.1007/s00792-019-01144-1

von Hegner I, Interplanetary transmissions of life in an evolutionary context. *Int. J. of Astrobio.*, 2020b; 19: 335-348.10.1017/S1473550420000099

Way MJ, Del Genio AD, Kiang NY, Sohl LE, Grinspoon DH, Aleinov I, Kelley M, Clune T. Was Venus the first habitable world of our solar system? *Geoph Res Lett* 2016;43(16) 8376-8383.

Way MJ, Del Genio ADD. Venusian habitable climate scenarios: Modeling Venus through time and applications to slowly rotating Venus-like exoplanets. *J Geophys Res Planets* 2020;125(5):e2019JE006276. doi.org/10.1029/2019JE006276

Whitman WB, Coleman DC, Wiebe WJ. Prokaryotes: The unseen majority. *P Natl Acad Sci* 1998;95(12): 6578–6583. doi.org/10.1073/pnas.95.12.6578

Wogan NF and Catling DC, When is chemical disequilibrium in Earth-like planetary atmospheres a biosignature versus an anti-biosignature? Disequilibria from dead to living worlds. The Astrophysical Journal 2020; 892 (2)127. doi.org/10.3847/1538-4357/ab7b81

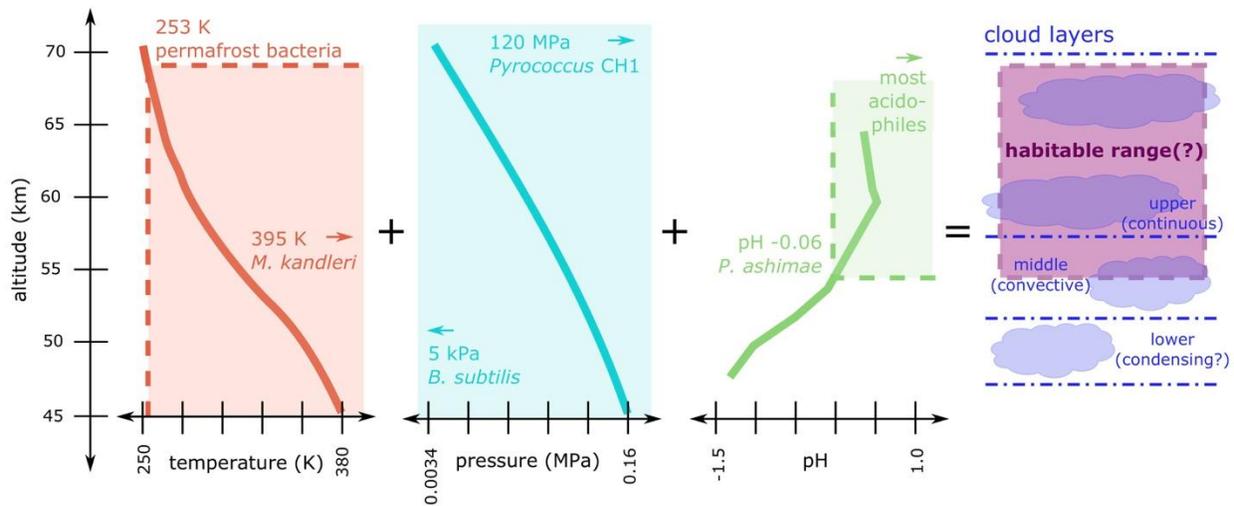

**Figure 1.** Three factors for which reasonable constraints exist that may point to favorable habitability conditions in the middle Venus atmosphere. Shown are the calculated temperature, pressure and pH prevailing in Venus cloud aerosols in the height range 45-70 km from the surface (solid lines), juxtaposed with the respective observed limitsfor terrestrial life (dashed lines and solid fill). Future missions may help to constrain additional major habitability variables such as water availability and ultraviolet radiation flux in this altitude range (see also Arking, 1996). The "habitable range" shown here for Venus is thus conservatively based on the "inhabited" range for Earth; the limits of terrestrial organisms may or may not reflect the possibilities for Venus.

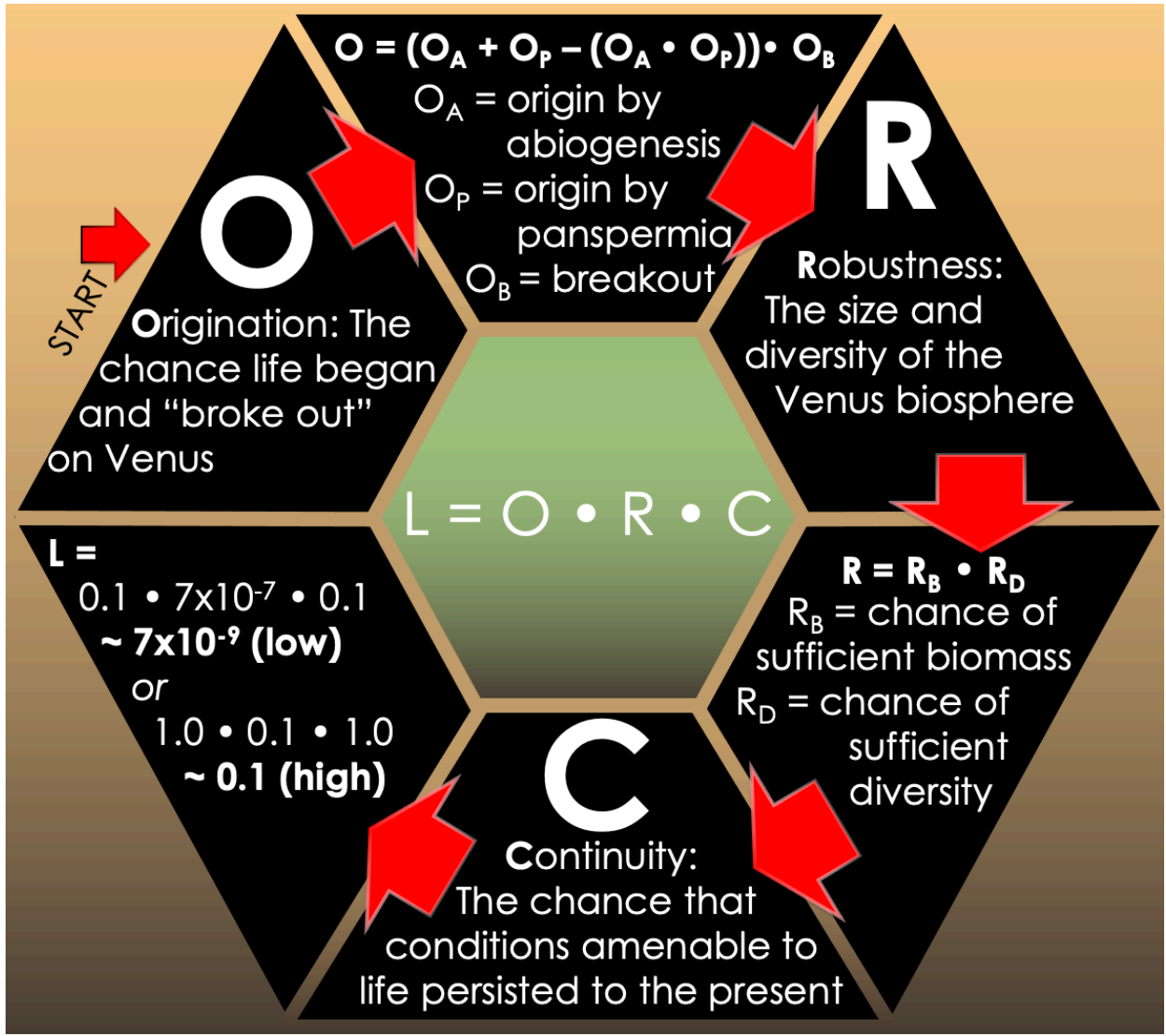

**Figure 2.** Schematic of the Venus Life Equation. Variables and equations elaborated in text. The final estimate for L in this figure represents an illustrative example using potential low and high ranges of estimates for each factor.

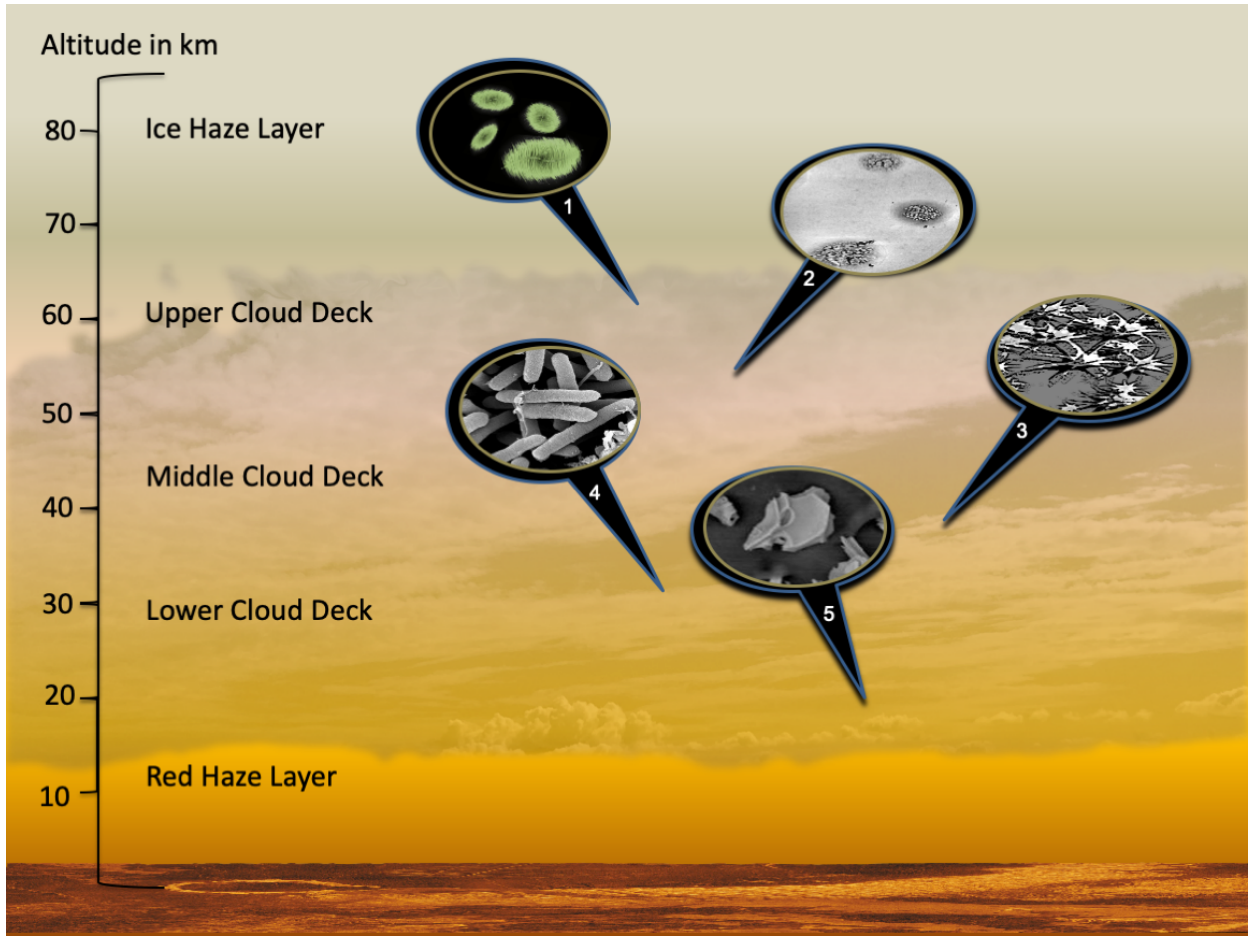

**Figure 3.** Notional particles potentially to be encountered in the Venus cloud decks, inspired by terrestrial atmospheric sampling, to guide future instrument and analysis selection: 1) Complex shapes with fluorescent properties, 2) Particulate aggregates of sulfates and related compounds, 3) Unidentified group of complex shapes adhered to an aerosol particle, 4) Objects that resemble Earth bacteria or archaea, and 5) volcanic ash particles.

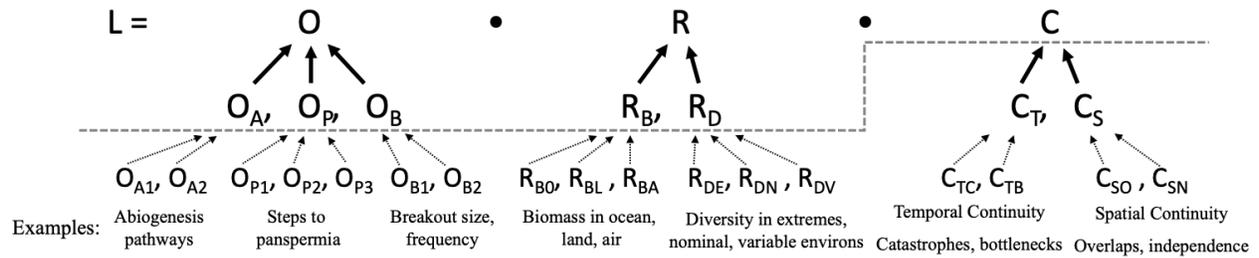

**Figure 4.** This illustration shows the breakdown of the Venus Life Equation into subfactors and how each of those subfactors can in turn be further broken down into finer details. The dotted line indicates the high level and global scale at which this paper addresses the equation. The examples of sub-sub factors are beyond the scope of this paper, and show only a representative, rather than exhaustive set of such factors. The contributors to each subfactor can be individual experiments, in-situ observations, individual biomes, geologic eras (such as the history of water loss), or models that may improve quantitative understanding. Better constraints or hard data on any sub factor in turn may improve the estimation of the overarching global factors of the VLE.